\documentclass{aa}
\usepackage{graphicx}
\usepackage{txfonts}
\usepackage{natbib}
\bibpunct{(}{)}{;}{a}{}{,}
\begin{document}
\title{Differential rotation in rapidly rotating F-stars
  \thanks{Table\,A1 is only available at the CDS via anonymous
    ftp to cdsarc.u-strasbg.fr (130.79.125.5) or via
    http://cdsweb.u-strasbg.fr/Abstract.html}
  \fnmsep 
  \thanks{Based on observations collected at the European Southern
    Observatory, La Silla, 69.D-0015(B)}}

%\titlerunning{}

\author{A. Reiners\inst{1}\thanks{Visiting Astronomer, German-Spanish Astronomical Centre,
    Calar Alto, operated by the Max-Planck-Institute for Astronomy,
    Heidelberg, jointly with the Spanish National Commission for
    Astronomy.}
        \and
        J.H.M.M. Schmitt\inst{1}}

\offprints{A. Reiners}

\institute{Hamburger Sternwarte, Universit\"at Hamburg,
  Gojenbergsweg 112, 21029 Hamburg, Germany\\
  \email{areiners, jschmitt@hs.uni-hamburg.de}}

\date{\today}

\abstract{We obtained high quality spectra of 135 stars of spectral
  types F and later and derived ``overall'' broadening functions in
  selected wavelength regions utilizing a Least Squares Deconvolution
  (LSD) procedure. Precision values of the projected rotational
  velocity $v\,\sin{i}$ were derived from the first zero of the
  Fourier transformed profiles and the shapes of the profiles were
  analyzed for effects of differential rotation.  The broadening
  profiles of 70 stars rotating faster than $v\,\sin{i} =
  45$\,km\,s$^{-1}$ show no indications of multiplicity nor of
  spottedness. In those profiles we used the ratio of the first two
  zeros of the Fourier transform $q_2/q_1$ to search for deviations
  from rigid rotation.  In the vast majority the profiles were found
  to be consistent with rigid rotation.  Five stars were found to have
  flat profiles probably due to cool polar caps, in three stars cuspy
  profiles were found.  Two out of those three cases may be due to
  extremely rapid rotation seen pole on, only in one case ($v\,\sin{i}
  = 52$\,km\,s$^{-1}$) solar-like differential rotation is the most
  plausible explanation for the observed profile.  These results
  indicate that the strength of differential rotation diminishes in
  stars rotating as rapidly as $v\,\sin{i} \ga
  50$\,km\,s$^{-1}$.}

\maketitle

\section{Introduction}
A magnetic dynamo is thought to lie at the root of all the activity
phenomena observed on the Sun and the stars.  Dynamo action requires
turbulent velocity fields and -- at least for the variety of
$\Omega$-dynamos -- differential rotation.  On the Sun surface
differential rotation is readily observed from the relative motion of
Sun spots and can be conveniently expressed in the form
\begin{equation}
  \label{DiffRotLaw}
  \Omega(l) = \Omega_{\rm Equator} (1 - \alpha \sin^2{l}),
\end{equation}
where $l$ denotes heliographic latitude and $\Omega_{\rm Equator}$ the
angular rotation speed at the equator.  Differential rotation is
described by the dimensionless parameter $\alpha$, which can be
thought of the difference of the equatorial and polar rotation
velocities relative to the equatorial velocity.

The measurement of rather subtle effects of differential rotation for
stars is quite difficult.  In principle three methods exist to
determine differential rotation in stars, first, by identifying
individual features on Doppler maps and following their surface
migration with time, by studying the rotation law with time, and by
studying line profiles obtained from high resolution data with very
good signal-to-noise ratio.  Using the latter approach,
\cite{Reiners03} were able to demonstrate the existence of essentially
a solar-like differential rotation law in 10 stars out of a sample
with 142 objects which were all rotating at intermediate velocities.
Note that only 32 stars had symmetric profiles amenable for a
sensitive search for differential rotation, and further, in slow
rotators the line profile changes caused by differential rotation
cannot be disentangled form other line broadening mechanisms, so that
the true percentage of differential rotators may in fact be higher.

The $\alpha$-parameter determined by \cite{Reiners03} for the 10 stars
in their sample indicates solar-like differential rotation.  These
results were surprising in that they seem to contradict the
differential rotation found through star spot migrations in some fast
rotators like, e.g., AB~Dor \citep{Donati97}.  In those stars the
effective $\alpha$ is very small and consequently the deviations from
rigid rotation are very small.  On the other hand, none of the stars
investigated by \cite{Reiners03} had $v\,\sin{i}$-values and
presumably rotation periods coming close to AB~Dor.  The line profile
analysis performed by \cite{Reiners03} can be applied to more rapid
rotators and in principle the effects of differential rotation can be
detected more easily since they are spread out over the line profile.
Yet in practice very rapid rotators pose extreme difficulties since,
first, their profiles are very shallow necessitating spectra with very
good signal, and second, line blending becomes severe.  In solar-like
stars rotating with $v\,\sin{i} > 100$\,km\,s$^{-1}$ or more, no
isolated spectral lines can be found and a suitable broadened line
profile can only be constructed with a robust deblending algorithm.
Such an algorithm has been developed by us and it is the purpose of
this paper to apply this algorithm to very rapid rotators, determine
their line broadening profiles, search for differential rotation and
compare the rotational properties of these rapid rotators to those of
the more slowly rotating stars studied by us previously.  In
Sect.\,\ref{Observations} we present our new observations, in
Sect.\,\ref{Deconvolution} we describe our deconvolution algorithm,
and in Sect.\,\ref{DistortProfiles} we discuss the obtained distorted
profiles.  New $v\,\sin{i}$-values are presented in
Sect.\,\ref{sect:vsini}, and our measurements of $q_2/q_1$, the ratios
of the zeros of second and first side lobe of the side lobes of the
Fourier transformed line profiles, in Sect.\,\ref{Results}.  The stars
with low $q_2/q_1$ are individually discussed, a summary of our paper
is given in Sect.\,\ref{Summary}.
   
\section{Observations}
\label{Observations}

\begin{figure*}[!ht]

  \includegraphics[height=\hsize,width=7cm,angle=-90,clip=]{0255.f1}

  \vspace{-3.7cm}

  \hspace{2.8cm}\includegraphics[height=5.2cm,width=2.5cm,angle=-90,clip=]{0255.f1i}
  \vspace{1.8cm}
  \caption{\label{normalfit}Data (thick line) and achieved fit (thin line) of
    HD~70\,958. Fit quality is typical for the whole wavelength
    region. The deconvolved ``overall'' profile is plotted in the
    inset. }

  \includegraphics[height=\hsize,width=7cm,angle=-90,clip=]{0255.f2}

  \vspace{-3.7cm}
  
  \hspace{2.8cm}\includegraphics[height=5.2cm,width=2.5cm,angle=-90,clip=]{0255.f2i}
  \vspace{1.8cm}
  \caption{\label{triples}Same as Fig.\ref{normalfit} but for the spectroscopic triple HD~118\,261.}
\end{figure*}

\begin{table}
  \caption[]{Observations}
  \label{tab:Observations}
  \begin{tabular}{cccc}
    \hline
    \hline
    \noalign{\smallskip}
    Date & \# Objects & Spectrograph & Resolution\\
    \noalign{\smallskip}
    \hline
    \noalign{\smallskip}
    24.02.--05.03.2002 & 69 & \textsc{Foces} & 40\,000 \\
    04.04.--05.04.2002 & 70 & \textsc{Feros} & 48\,000 \\
    \noalign{\smallskip}
    \hline
  \end{tabular}
\end{table}

In our study of rapidly rotating stars we used two spectrographs
covering the full optical wavelength region. For the southern
hemisphere objects we used \textsc{Feros} at ESO (La Silla, Chile), for the
northern hemisphere we used \textsc{Foces} at the DSAZ (Calar Alto, Spain). The
resolving power of both spectrographs is $R \sim 48\,000$ and $R \sim
40\,000$, respectively. Our sample consists of 135 stars of spectral
types F and later. Details of the observing runs are listed in
Table\,\ref{tab:Observations}; 69 stars were observed with \textsc{Foces}, 70
stars with \textsc{Feros}, and four objects were observed
during both campaigns. We focused on moderately fast rotators with
literature values of $v\,\sin{i} > 45$\,km\,s$^{-1}$ since the
resolution of both instruments is sufficient to study such profiles.  Nine
stars without any published measurements of $v\,\sin{i}$ were observed in
addition.  To guarantee high quality spectra we observed only
stars brighter than $m_{\rm V} = 6.5$\,mag; all spectra have
signal-to-noise ratios in excess of 300.

\section{The deconvolution process}
\label{Deconvolution}

Spectral lines of fast rotating F-type stars are severely blended, the optical
spectrum of an F-type star with  $v\,\sin{i} >$ 40\,km\,s$^{-1}$
contains hardly a single isolated absorption line. Thus we did not use
single absorption lines, but rather derived an ``overall'' broadening function
in a Least Squares Deconvolution process. Our data cover the
wavelength region between 3800 and 8200\,\AA. Substantial parts of the
observed spectral region are not useful for deriving information about stellar
rotation since light-ion lines of hydrogen or helium are pressure
dominated, and further, large spectral regions and the strong lines of Ca or Na
are contaminated or even dominated by telluric features. Thus we
searched for spectral regions with a significant number of
rotationally dominated adequately spaced lines.

\begin{table}
  \caption{\label{tab:regions}The three wavelength regions used in the 
         deconvolution process and the approximate number of lines used}
  \begin{tabular}{cccc}
    \hline
    \hline
    \noalign{\smallskip}
    \# & Wavelength region & coverage & approx. \# of lines\\
    \noalign{\smallskip}
    \hline
    \noalign{\smallskip}
    1 & 5440 -- 5880 \AA & 440 \AA & 300 \\
    2 & 6040 -- 6270 \AA & 230 \AA & 150 \\
    3 & 6330 -- 6450 \AA & 120 \AA & 50 \\
    \noalign{\smallskip}
    \hline
  \end{tabular}
\end{table}

We identified three such wavelength regions given in
Table\,\ref{tab:regions} and derived broadening profiles in each
region independently. Since the theoretical line depths match the
observational ones only poorly, the equivalent widths of the
incorporated lines were optimized as well. The iterative procedure was
started by constructing a delta-template according to stellar
temperature and atomic data taken from the Vienna Atomic Line Database
\citep{VALD}. A first-guess broadening profile with each pixel as a
free parameter was deconvolved from the spectrum leaving the template
at first fixed. In a second step the equivalent widths of the template
were optimized without changing the broadening profile. For the
construction of the broadening profile a flat continuum normalization
of the spectrum obtained by eye was used.  After the first iteration
of the line broadening profile and template a continuum correction was
determined by allowing the continuum to vary as a smooth spline and
yield best agreement between the preliminary fit and observed
spectrum. These continuum corrections are typically very small ($<
1\%$). With the improved continuum estimate the broadening profile and
the equivalent widths were further optimized iteratively.  No
addtional continuum corrections were found to be necessary.  Using
this technique the spectral lines are effectively deblended, the
information contained in all spectral lines is used and the
signal-to-noise ratio is significantly enhanced. The consistency of
the final fit is checked by comparing theoretical line depths to the
derived ones. To demonstrate the quality of our fits we show in
Figs.\,\ref{normalfit} and \ref{triples} our model spectrum (thin
lines) in comparison to the data (thick lines) in a small wavelength
region for two examples (HD~70\,958, HD~118\,261). The derived
profiles are shown in the insets.  HD~70\,958 (Figs\,\ref{normalfit})
is a typical fast rotator in our sample, the crowded spectrum of
HD~118\,261 (Fig.\,\ref{triples}) turns out to be produced by a
spectroscopic triple.

\begin{table}
  \caption{\label{ObservedTwice}Targets observed in both observing campaigns. 
              Note that the minimum error of $v\,\sin{i}$ is adopted as 5\%
              (cp. Sect.\,\ref{sect:vsini}).}
  \begin{tabular}{ccccc}
    \hline
    \hline
    \noalign{\smallskip}
    & \multicolumn{2}{c}{\textsc{Feros} spectra} & \multicolumn{2}{c}{\textsc{Foces} spectra} \\ 
    HD & $v\,\sin{i}$ & $q_2/q_1$ & $v\,\sin{i}$ & $q_2/q_1$ \\
    \noalign{\smallskip}
    \hline
    \noalign{\smallskip}
    103\,313 & $66.6 \pm 3.3$ && $67.4 \pm 3.3$ &\\
    124\,425 & $23.5 \pm 4.0$ && $23.5 \pm 4.1$ &\\
    125\,451 & $40.5 \pm 2.0$ && $40.7 \pm 2.0$ &\\
    147\,449 & $77.1 \pm 3.9$ & $1.80 \pm 0.03$ & $76.4 \pm 3.9$ & $1.79 \pm 0.04$\\
    \noalign{\smallskip}
    \hline
  \end{tabular}
\end{table}

For the sake of homogeneity we used the broadening profiles
deconvolved from region~1 in Table\,\ref{tab:regions} for our
analysis.  This region contains more lines than all others and the
deconvolution process turned out to be more robust than in the regions
containing fewer lines.  We did compare the profiles derived from
regions~2 and 3 to those derived from region~1 and found them to be
fully consistent whenever profiles could be derived from regions~2 and
3. This of course implies that the zeros of the Fourier transforms
used in the following are consistent, too. Note, however, that in a
few cases of very rapid rotators no convergent solutions could be
derived from regions~2 and 3.

Four target stars were observed during both observing campaigns and
broadening profiles were independently derived for them. Consistent
profiles could be found for all data sets. Independently derived
results discussed in the following sections are given in
Table\,\ref{ObservedTwice}.

\section{Distorted profiles}
\label{DistortProfiles}

The high quality of the derived line profiles allows us to directly
identify the mechanisms influencing the shape of absorption lines.
Binaries and even triples (cf., Fig.\,\ref{triples}) can be found as
long as the spectral types do not differ too much. The broadening
profile of a (differentially) rotating single star is expected to be
symmetric; asymmetric velocity fields such as turbulent flows are at
least an order of magnitude smaller than the rotational effects
studied in this work. Therefore significantly asymmetric profiles must
be due to other mechanisms such as multiplicity or spottedness.

Among our sample of 135 stars we found two profiles with three
clearly isolated components -- one of them is shown in
Fig.\,\ref{triples}. Three spectroscopic binaries with separated
components were found.  16 profiles are likely due to double or triple
systems with blended profiles of the different components.  All
multiple stars are listed in Table\,\ref{tab:multiples} with the values
of $v\,\sin{i}$ of their components derived as described in
Sect.\,\ref{sect:vsini}.

\begin{table}
  \caption{\label{tab:multiples}Spectroscopic multiples and $v\,\sin{i}$ of their components}
  \begin{tabular}{rrrr}
    \hline
    \hline
    \noalign{\smallskip}
    HD & $v\,\sin{i}_{1}$ & $v\,\sin{i}_{2}$ & $v\,\sin{i}_{3}$\\ 
    &[km\,s$^{-1}$]&[km\,s$^{-1}$]&[km\,s$^{-1}$]\\
    \noalign{\smallskip}
    \hline
    \noalign{\smallskip}
    \multicolumn{4}{c}{\emph{triples with separated components}}\\
    58728 & $29.3 \pm 1.5$& $11.0 \pm 1.0$ & $30.9 \pm 2.2$\\
    118261 & $4.7 \pm 3.6$ & $12.6 \pm 1.0$ & $5.2 \pm 1.0$\\
    \noalign{\smallskip}
    \hline
    \noalign{\smallskip}
    \multicolumn{4}{c}{\emph{doubles with separated components}}\\
    99453 & $4.9 \pm 1.2$ & $<4.4$ &\\
    114371  & $4.6 \pm 4.0$ & $<4.4$ &\\
    177171  & $37.5 \pm 1.9$ & $50.0 \pm 3.0$ &\\
    \noalign{\smallskip}
    \hline
    \noalign{\smallskip}
    \multicolumn{4}{c}{\emph{doubles or triples with blended components}}\\
    12230  & $206.0 \pm 5.0$ & $44.0 \pm 4.0$ &\\
    27176  & $103.0 \pm 4.0$ & $6.9 \pm 1.7$ &\\
    28052  & $243.0 \pm 5.0$ & $28.0 \pm 5.0$ & $22.0 \pm 5.0$\\
    50635  & $154.0 \pm 5.0$ & $<6.0$ &\\
    51733  & $103.0 \pm 5.0$ & $<6.0$ &\\
    65925  & $72.9 \pm 3.6$ & $<6.0$ &\\
    80671  & $25.7 \pm 2.1$ & $<6.0$ &\\
    82434  & $158.0 \pm 10.0$ & $16.0 \pm 3.0$ &\\
    87500  & $191.0 \pm 10.0$ & $10.3 \pm 2.0$ &\\
    89025  & $84.0 \pm 8.0$ & $19.0 \pm 8.0$ & $11.4 \pm 2.0$ \\
    109585  & $91.4 \pm 5.0$ & $<6.0$ &\\
    113139  & $91.1 \pm 5.0$ & $<6.0$ &\\
    114435  & $80.4 \pm 5.0$ & $<6.0$ &\\
    125442  & $153.0 \pm 8.0$ & $48.2 \pm 3.0$ &\\
    144415  & $100.6 \pm 5.0$ & $<6.0$ &\\
    153580  & $45.7 \pm 5.0$ & $7.0 \pm 3.0$ &\\
    \noalign{\smallskip}
    \hline
  \end{tabular}
\end{table}

Unlike the multiple stars listed in Table\,\ref{tab:multiples}, 23
clearly asymmetric profiles were found which are obviously not due to
multiple components.  These profile asymmetries might be caused by
star spots or pulsations. In Fig.\,\ref{plot:HRD} these stars are
plotted with crosses in an HR-diagram, and we will discuss this
diagram in Sect.\,\ref{sect:HRD}. A shape flag is given to the
profiles in the last column of Table\,\ref{tab:vsini}.

\section{New values of v\,sini}
\label{sect:vsini}

\begin{figure}
  \centering
  \resizebox{\hsize}{!}{\includegraphics[angle=-90,clip=]{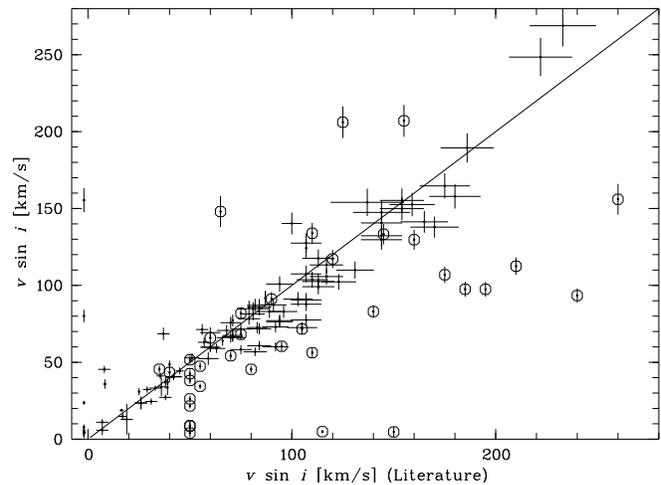}}
  \caption{\label{plot:vsini} Values of $v\,\sin{i}$ from this work vs.
    literature values. Those found only in \cite{Uesugi82} are marked
    with an open circle.}
\end{figure}

The projected rotational velocity $v\,\sin{i}$ was measured from the
first zero $q_{1}$ of the Fourier transform. The error of $v\,\sin{i}$
directly depends on the quality of the deconvolved profile and on the
precision of a determination of $q_{1}$. In most cases this intrinsic
precision is of the order of 1\,km\,s$^{-1}$. The reliability of the
deconvolved profiles was tested with synthetic data. Our tests
showed that the systematic errors of the determined values of
$v\,\sin{i}$ induced by the deconvolution process are always less than
$5\,\%$. In Table\,\ref{tab:vsini} we list our the results; as effective
error we adopted the maximum of the error of determining $q_1$ and of
deconvolving the profile (5\,\%).

A comparison of our results with data from the literature is presented in
Fig.\,\ref{plot:vsini}. Literature values are taken from the
comprehensive catalog of \cite{Glebocki00} and from \cite{Royer02}.
Values only found in the catalog of \cite{Uesugi82} -- which is
contained in \cite{Glebocki00} -- are indicated with open circles. For
nine of our 135 objects no measurements of $v\,\sin{i}$ are reported
in \cite{Glebocki00} nor are contained in \cite{Royer02}.  These
values are plotted at $v\,\sin{i}_{\rm Literature} =
-2.0$\,km\,s$^{-1}$ in Fig.\,\ref{plot:vsini}.  While the consistency
of our results with measurements made by other authors is quite good, the values
taken from \cite{Uesugi82} show very large discrepancies and cannot be
considered reliable.  The projected rotational velocities of 37 stars are
only reported in \cite{Uesugi82} and we suggest to use our redetermined values.

\section{The profile's shape and the value of $\mathbf{q_{2}/q_{1}}$}
\label{Results}

\begin{figure*}[ht]
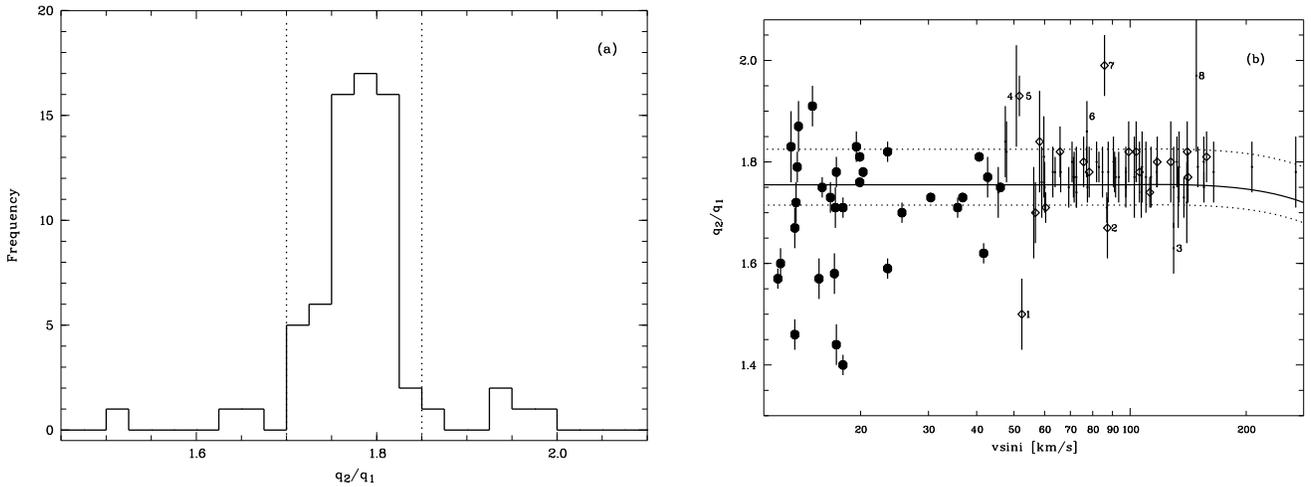

  \resizebox{.5\hsize}{!}{\includegraphics[angle=-90,clip=]{0255.f4a}}
  \resizebox{.5\hsize}{!}{\includegraphics[angle=-90,clip=]{0255.f4b}}
  \caption{\label{plot:q2q1_vsini}(a) Distribution of the ratio $q_2/q_1$
    in our sample; (b) $q_2/q_1$ plotted vs. $v\,\sin{i}$. Dotted
    lines mark the region of $q_2/q_1$ typical for solid body rotation
    with linear limb darkening and arbitrary limb darkening
    coefficient $\epsilon$. The solid line in (b) marks the value
    expected for rigid rotation and a linear limb darkening
    coefficient $\epsilon = 0.6$.  Stars denoted with numbers in (b)
    are given in Table\,\ref{tab:smallq2q1} (dots: \textsc{Feros}
    data; rhombs: \textsc{Foces} data); filled circles: data from
    \cite{Reiners03}.}
\end{figure*}

\begin{table*}
  \caption{\label{tab:smallq2q1}Parameters of the stars with $q_2/q_1 < 1.72$
               or $q_2/q_1 > 1.83$ and significantly different from 1.76. Numbers
               as in Fig.\,\ref{plot:q2q1_vsini}b (see text).}
  \begin{tabular}{c|rlccrcccc}
    \hline
    \hline
    \noalign{\smallskip}
    No. & HD & Type & $T_{\rm eff}$ &$M_{\rm V}$ & $v\,\sin{i}$ & $q_{2}/q_{1}$ & $\alpha$ & $v_{\rm e, rigid}$ & $i$  \\
    (Fig.\,\ref{plot:q2q1_vsini})&&&[K]&[mag]&[km\,s$^{-1}$]&&&[km\,s$^{-1}$]\\    \noalign{\smallskip}
    \hline
    \noalign{\smallskip}
    1 & 67483 & F3V & 6209 & 2.07 & $52.4 \pm 4.1$ & $1.50 \pm 0.07$ &$0.33 \pm 0.13$& (580) & (5\degr)\\
    \noalign{\smallskip}
    \hline
    \noalign{\smallskip}
    2 & 182640 & F1IV-V & 7016 & 2.46 & $87.3 \pm 4.4$ & $1.67 \pm 0.06$ &$0.18 \pm 0.12$& 380 & 13\degr\\
    3 & 82554 & F3IV & 6272 & 1.32 & $129.7 \pm 6.5$ & $1.63 \pm 0.05$ &$0.12 \pm 0.12$& 435 & 17\degr\\
    \noalign{\smallskip}
    \hline
    \noalign{\smallskip}
    4 & 55892 & F0IV & 6802 & 2.37 & $50.7 \pm 2.5$ & $1.93 \pm 0.10$ & \multicolumn{3}{c}{\emph{variable star}}\\
    5 & 90589 & F2IV & 6794 & 2.95 & $51.6 \pm 2.6$ & $1.93 \pm 0.04$ & \multicolumn{3}{c}{\emph{variable star}}\\
    6 & 106022 & F5V & 6651 & 2.28 & $77.3 \pm 3.9$ & $1.86 \pm 0.06$ & \multicolumn{3}{c}{\emph{star in double system}}\\
    7 & 138917 & F0IV & 7442 & 1.53 & $85.8 \pm 5.4$ & $1.99 \pm 0.06$ & \multicolumn{3}{c}{\emph{star in double system}}\\
    8 & 175813 & F2V & 6659 & 1.99 & $148.5 \pm 8.0$ & $1.97 \pm 0.15$ & \multicolumn{3}{c}{\emph{W UMa type star}}\\
    \noalign{\smallskip}
    \hline
  \end{tabular}
\end{table*}

The shape of a rotationally broadened line profile can conveniently be
described in Fourier space. The zeros of a Fourier transformed line
profile contain information on the projected rotational velocity
$v\,\sin{i}$, and differential rotation.  Specifically,
\cite{Reiners02} show that the ratio of the second and the first zeros
$q_2/q_1$ is a reliable observable of solar-like differential
rotation. Regardless of the value of the limb darkening coefficient
$\epsilon$, the ratio $q_2/q_1$ always stays between 1.72 and 1.83 in
a rigidly rotating star with a linear limb darkening law, and one
finds $q_2/q_1 = 1.76$ for a limb darkening coefficient of $\epsilon =
0.6$.  Solar-like differential rotation with the equator rotating
faster than the pole diminishes the ratio $q_2/q_1$, and values of
$q_2/q_1 < 1.72$ are found in stars with solar-like differential
rotation stronger than 10\,\% (i.e., $\alpha > 0.10$ in
Eq.\,\ref{DiffRotLaw}).

In order to measure the ratio $q_2/q_1$ the data quality must allow to
follow the Fourier transformed broadening profile to at least the
second sidelobe to obtain a reliable measurement of the second zero
$q_2$. For our data this is the case for all symmetric profiles of
stars rotating faster than $v\,\sin{i} = 45$\,km\,s$^{-1}$.  In 21
cases our new determinations of $v\,\sin{i}$ yielded values lower than
this threshold; note that ten of these stars were supposed to rotate
faster according to \cite{Uesugi82}.

Disregarding the 44 distorted profiles and 21 profiles of stars
rotating too slowly, we were able to measure the ratio $q_2/q_1$ for
70 of our sample stars. A histogram of the distribution of $q_2/q_1$
among these 70 targets is shown in Fig.\,\ref{plot:q2q1_vsini}a. For
62 stars (89\,\%) the values of $q_2/q_1$ are consistent with solid
body rotation, five stars (7\,\%) have values of $q_2/q_1 > 1.83$ and
for three cases (4\,\%) $q_2/q_1 < 1.72$ was found; in one of these
three cases (HD~82\,554) the measured $q_2/q_1$-ratio is consistent
with solid body rotation to within the 1-$\sigma$ errors.

In Fig.\,\ref{plot:q2q1_vsini}b we plot the values of $q_2/q_1$ vs.
$v\,\sin{i}$ together with results from \cite{Reiners03}.  The solid
line indicates the value expected for rigid rotators with a linear
limb darkening parameter $\epsilon = 0.6$, dashed lines mark the
region calculated for extreme limb darkening $\epsilon = 0.0$ and
$\epsilon = 1.0$. Since the limb darkening law is only poorly known
over the whole spectral region, values between the dashed lines are
considered consistent with solid body rotation. Note that according to
\cite{Reiners03a} values of $q_2/q_1 < 1.76$ are expected for very
fast rotators; this is the cause for the bending of the solid and
dashed lines for high $v\,\sin{i}$-values
(Fig.\,\ref{plot:q2q1_vsini}), which will be further discussed in
Sect.\,\ref{lowq2q1}. The stars with values of $q_2/q_1$ outside the
region marked by the dashed lines and significantly different from
1.76 are indicated with numbers in Fig.\,\ref{plot:q2q1_vsini}b and
are given in Table\,\ref{tab:smallq2q1}, they are discussed in the
following.

\subsection{Stars with large values of $\mathsf{q_2/q_1}$}

Stars with a value of $q_2/q_1 > 1.76$ have trough-like profiles
flattened in the center. Slowly rotating surface elements are
apparently emitting less flux than expected. This could be either due
to surface elements with small projected velocities emitting less flux
or very few surface elements rotating that slowly. According to
\cite{Reiners02, Reiners03} this is possible in cases of cool polar
spots or anti solar-like differential rotation with the pole rotating
faster than the equator. Since no evidence exists for the latter, cool
polar caps are considered the most probable explanation for the five
stars with $q_2/q_1 > 1.76$. Each of these stars is either a member of
a double system or classified as variable in the literature as given
in Table\,\ref{tab:smallq2q1}. We suspect that additional mechanisms
influence the surface structure of these stars, but a more detailed
discussion goes beyond the scope of this paper.

\subsection{Stars with small values of $\mathsf{q_2/q_1}$}
\label{lowq2q1}
Three of the 70 stars of our sample have values significantly smaller
than $q_2/q_1 = 1.76$. Solar-like differential rotation is the only
mechanism known that sufficiently reduces the value of $q_2/q_1$ in
slow rotators. The strength of differential rotation in terms of
$\alpha$ in Eq.\,\ref{DiffRotLaw} can be calculated from the ratio
$q_2/q_1$, and \cite{Reiners03} derived differential rotation in ten
F-type stars with values of $v\,\sin{i} < 50$\,km\,s$^{-1}$.

\begin{figure}
  \centering
  \resizebox{\hsize}{!}{\includegraphics[angle=-90,clip=]{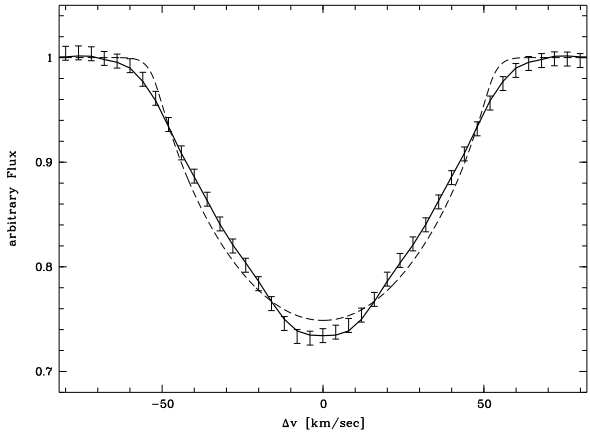}}
  \caption{\label{plot:HD67483} Derived ``overall'' broadening profile of
    the suspected differential rotator HD~67\,483. We plot the
    symmetrized profile (solid line) over the original error bars to
    demonstrate the profile's symmetry. For comparison a profile of a
    rigidly rotating star with the same value of $v\,\sin{i}$ is
    overplotted (dashed line).}
\end{figure}

\cite{Reiners03a} shows that gravitational darkening can also diminish
the value of $q_2/q_1$ in very rapidly rotating stars. In that case
the reduction in $q_2/q_1$ depends on the equatorial velocity $v_{\rm
  e}$, rather than the projected velocity $v\,\sin{i}$. Thus for
extremely rapid rotators (perhaps viewed under small inclination
angles), values of $q_2/q_1 <1.72$ are possible even for rigid
rotation.  Assuming that the derived broadening profile is not
dominated by the intrinsic variations of the local profiles due to
temperature and gravity, it is possible to calculate $v_{\rm e}$ from
the value of $q_2/q_1$ \citep[cf][]{Reiners03a}. The required values
of differential rotation $\alpha$ and, following the idea of very
rapid solid body rotation, equatorial velocity $v_{\rm e}$ and
inclination angle $i$ are calculated for the three candidates. The
results are given in Table\,\ref{tab:smallq2q1}.  The small ratios
$q_2/q_1$ of HD~82\,554 and HD~182\,640 may thus be due to very rapid
solid body rotation, but the equatorial velocity calculated for the
profile of HD~67\,483 is larger than break up velocity. Therefore
differential rotation is the most probable explanation for the
peculiar shape of the line profiles of HD~67\,483. In
Fig.\,\ref{plot:HD67483} the profile of HD~67\,483 is shown.

\begin{figure}
  \centering
  \resizebox{\hsize}{!}{\includegraphics[angle=-90,clip=]{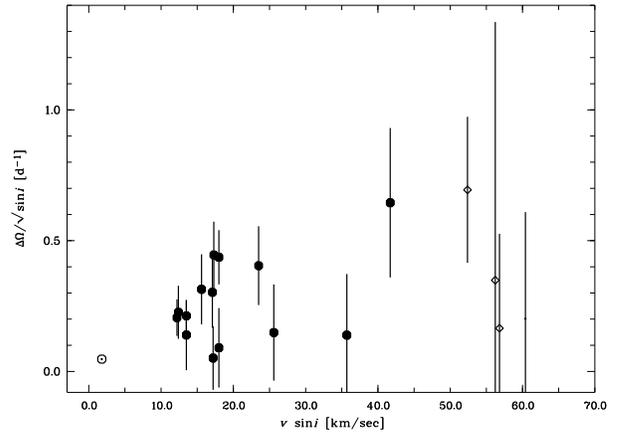}}
  \caption{\label{plot:DOmega_vsini} Equator-pole beat frequency
    $\Delta \Omega$ plotted vs. $v\,\sin{i}$ for stars with values of
    $q_2/q_1 < 1.72$ which are likely due to solar-like differential
    rotation. Symbols as in Fig.\,\ref{plot:q2q1_vsini}b. Note that
    HD~82\,554 and HD~182\,640 are not plotted since their small
    values of $q_2/q_1$ can also be due to gravitational darkening.}
\end{figure}

In Fig.\,\ref{plot:DOmega_vsini} we show the beat frequency $\Delta
\Omega$ between equator and pole vs.  $v\,\sin{i}$ for those stars
with detected signatures of differential rotation; we also included
more slowly rotating stars discussed by \cite{Reiners03}, the value of
solar differential rotation has also been included.
Fig.\,\ref{plot:DOmega_vsini} suggests an increase in beat frequency
up to rotational velocities of about 40\,km\,s$^{-1}$ followed by a
sharp decrease in $\Delta \Omega$ for higher velocities. Note that
this latter interpretation depends on the interpretation of the two
stars HD~82\,554 and HD~182\,640 as solid body vs. differential
rotators. If the derived values of $q_2/q_1$ are due to differential
rotation rather gravitational darkening, similar beat frequencies of
about $\Delta \Omega \approx 0.6$ would be found for these two
extremely rapid rotators.

\subsection{Location in the HR-diagram}

\begin{figure}
  \centering
  \resizebox{\hsize}{!}{\includegraphics[angle=-90,clip=]{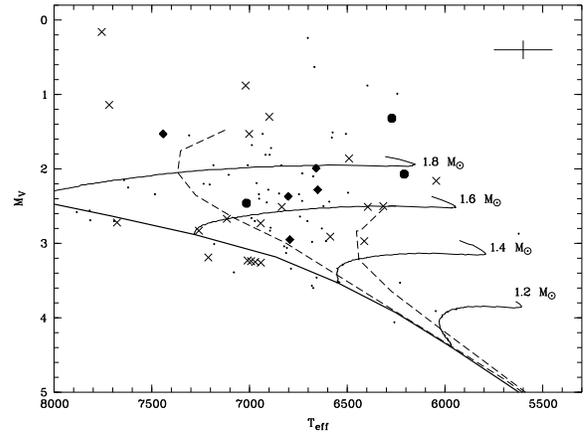}}
  \caption{\label{plot:HRD}HR-diagram of the suspected single stars
    of our sample. Different symbols indicate profile shape: crosses:
    asymmetric; circles: $q_2/q_1 < 1.72$ (cuspy); rhombs: $q_2/q_1 >
    1.83$ (flat); points: ``regular'' rotational broadening.
    Evolutionary tracks are from \cite{Siess00}. }
\end{figure}

\label{sect:HRD}
In Fig.\,\ref{plot:HRD} all suspected single stars in our sample
are plotted in an
HR-diagram with different symbols depicting their profile shapes' as
explained in the caption. Values $M_{\rm V}$ and $T_{\rm eff}$ are
calculated from $uvby\beta$ photometry using the program UVBYBETA
published by \citet{Moon85a}; data are taken from \citet{Hauck98}. For
$T_{\rm eff}$ a new calibration by \citet{Napiwotzki93} based on the
grids of \citet{Moon85b} was used, the statistical error of the
temperature determination is about $\Delta T_{\rm eff} = 150$\,K,
typical error bars are plotted in Fig.\,\ref{plot:HRD}. The zero-age
main sequence and evolutionary tracks according to \cite{Siess00} are
also shown.
Fig.\,\ref{plot:HRD} shows that no significant difference exists
between the distribution of the stars with different profile shapes.
``Regular'' broadening profiles, peculiar profiles and extremely flat
or cuspy shaped profiles do not show concentrations in any area of the
HR-diagram.

\section{Summary}
\label{Summary}

We performed high-resolution spectroscopy for a sample of 135 stars of
spectral types F and later, focusing on fast rotators with values of
$v\,\sin{i} > 45$\,km\,s$^{-1}$. Using a Least Squares Deconvolution
method ``overall'' broadening profiles were derived in a dedicated
spectral region containing typically about 300 absorption lines.
Broadening profiles were derived in two wavelength regions for
comparison, good consistency was found. We determined the values of
$v\,\sin{i}$ from the first zeros $q_1$ of the Fourier transformed
broadening profiles.  For nine stars no previous measurements were
reported in the literature, for 37 stars our measurements are far more
reliable than the previously reported measurements listed in catalog
of \cite{Uesugi82}.

The shape of the broadening profiles were scrutinized in wavelength
space.  We found three profiles with separated spectroscopic binaries,
two spectra with separated spectroscopic triples and 21 profiles that
are likely a sum of two (or possibly three) profiles indicating a
binary (or triple) system. For 23 stars peculiar and strongly
asymmetric profiles were found, probably caused by spots or
pulsations.

21 stars were found to have projected rotational velocities of
$v\,\sin{i} < 45$\,km\,s$^{-1}$; for those stars the spectral
resolution was insufficient to follow their Fourier transforms to the
second zero $q_2$.  For the remaining 70 stars the first two zeros of
the Fourier transforms could be measured and the ratio $q_2/q_1$ was
used as a reliable tracer of the profile shapes'. The vast majority of
these stars (89\,\%) were found to have profiles consistent with the
assumption of a rigidly rotating unspotted surface. Five stars (7\,\%)
have flattened profiles likely due to polar spots (or even anti
solar-like differential rotation), all of them are cited as multiples
or variables in the literature.

Three stars show cuspy broadening profiles inconsistent with solid
body rotation ($q_2/q_1 < 1.76)$ indicating solar-like differential
rotation.  Alternatively, in the cases of HD~82\,554 and HD~182\,640
very rapid rotation viewed under small inclination angles may be the
mechanism causing the peculiar shape. Thus the comparably small
deviation from the classical shape does not necessarily require
differential rotation. However, for the case of HD~67\,483 the
measured $q_2/q_1$-value of the profile ($q_2/q_1 = 1.50 \pm 0.07$)
would require a rotational velocity much larger than break-up velocity
to be consistent with solid body rotation. Therefore, differential
rotation with a value of $\alpha = 0.33 \pm 0.13$ is the most probable
explanation for the observed line profile.

Considering HD~82\,554 and HD~182\,640 as rigid rotators (note that
for them differential rotation with $\alpha \approx 0.2$ is possible
but not necessary), only one star with unambiguous indications for
solar-like differential rotation stronger than $\alpha = 0.1$ was
found in the sample.  \cite{Reiners03} found ten stars with
indications of significant solar-like differential rotation in their
sample of 32 ``slow'' rotators ($v\,\sin{i} < 50$\,km\,s$^{-1}$). Thus
regardless of whether HD~82\,554 and HD~182\,640 are treated as solid
of differential rotators we conclude that solar-like differential
rotation in terms of $\alpha$ diminishes in rapidly rotating stars.
With a rotational period of $P/\sin{i} = 1.5$\,d as derived from
$v\,\sin{i}$ and with $\alpha = 0.33$, the equator laps the pole once
every 5.7\,d in HD~67\,483 (in the Sun this is the case about every
130\,d). The results derived from our sample show this is to be one of
the most extreme cases of stellar differential rotation possible (if
HD~82\,554 and HD~182\,640 are differential rotators their lap time is
larger than nine days).

We conclude that in F-type stars strong differential rotation with
$\alpha \ge 0.1$ seems to be a common phenomenon in slow rotators
($v\,\sin{i} \la 50$\,km\,s$^{-1}$) but untypical in the more rapid
ones.  In that sense our results are consistent with those from
Doppler Imaging studies claiming values less $\alpha = 0.01$ for some
very rapid rotators \cite[e.g.,][]{Donati97}.

\begin{acknowledgements} 
  A.R. acknowledges financial support from Deutsche
  Forschungsgemeinschaft DFG-SCHM 1032/10-1. This work has made
  extensive use of the SIMBAD database at CDS. \\
  We thank our anonymous referee for a very helpful report taking less
  than a typical rotation period of our rapidly rotating stars.
\end{acknowledgements}

\appendix

\section{Table of projected rotational velocities.}
\label{Appendix}

Shape flags:\\
\begin{tabular}{lcl}
b & -- & {\bfseries b}lended multiple (suspected)\\
s & -- & {\bfseries s}potted star (suspected)\\
a & -- & {\bfseries a}symmetric profile (reason unidentified)\\
m2 & -- & {\bfseries m}ultiple with {\bfseries 2} separated components\\
m3 & -- & {\bfseries m}ultiple with {\bfseries 3} separated components\\[1mm]
c & -- & symmetric profile, but {\bfseries c}andidate for spottedness
\end{tabular}

\begin{table*}
  \caption[]{Projected rotational velocites of our sample stars.}
  \label{tab:vsini}
  \begin{tabular}{rrcrrrrrc}
    \hline
    \hline
    \noalign{\smallskip}
    HD & HR & Name & $B-V^{\rm a}$ & $v\,\sin{i}$  & $¸\Delta~v\,\sin{i}$ & $q_{\rm 2}/q_{\rm 1}$ & $\Delta q_{\rm 2}/q_{\rm 1}$ & Shape \\
    & & & & [km\,s$^{-1}$] & [km\,s$^{-1}$] & & & Flag\\
    \noalign{\smallskip}
    \hline
    \noalign{\smallskip}
     12230  &   581  &   47    Cas          &   0.34  &  206.0  &  10.3 &       &       & b  \\ 
 13480  &   642  &    6    Tri          &   0.77  &  35.8  &   3.1 &       &       & s  \\ 
 15524  &   728  &                      &   0.41  &  59.7  &   3.0 & 1.81 & 0.08 &    \\ 
 17904  &   855  &   20    Per          &   0.42  &  54.2  &   2.7 &       &       & a  \\ 
 24357  &  1201  &                      &   0.35  &  65.8  &   3.3 & 1.82 & 0.05 & c  \\ 
 25202  &  1238  &                      &   0.31  &  164.7  &   8.2 & 1.78 & 0.06 &    \\ 
 25570  &  1254  &                      &   0.37  &  34.3  &   1.7 &       &       &    \\ 
 25867  &  1269  &   42\,$\psi$ \,$\tau$          &   0.35  &  45.4  &   2.3 &       &       & a  \\ 
 27176  &  1331  &   51    \,$\tau$          &   0.27  &  100.8  &   5.0 &       &       & b  \\ 
 27459  &  1356  &   58    \,$\tau$          &   0.22  &  78.3  &   3.9 & 1.78 & 0.05 & c  \\ 
 28052  &  1394  &   71    \,$\tau$          &   0.26  &  248.4  &  12.4 &       &       & b  \\ 
 28271  &  1406  &                      &   0.54  &  33.2  &   1.7 &       &       & a  \\ 
 29391  &  1474  &   51    Eri          &   0.28  &  71.8  &   3.6 &       &       & s  \\ 
 29875  &  1502  &     \,$\alpha$ Cae          &   0.34  &  47.8  &   2.4 & 1.82 & 0.06 &    \\ 
 29992  &  1503  &     \,$\beta$ Cae          &   0.39  &  97.5  &   4.9 & 1.75 & 0.04 &    \\ 
 30034  &  1507  &                      &   0.25  &  103.6  &   5.2 & 1.82 & 0.06 & c  \\ 
 30912  &  1554  &                      &   0.36  &  207.0  &  10.4 & 1.79 & 0.05 &    \\ 
 32743  &  1649  &     \,$\eta$1\,$\pi$c          &   0.42  &  21.6  &   1.6 &       &       &    \\ 
 33262  &  1674  &     \,$\zeta$ Dor          &   0.52  &  14.8  &   2.1 &       &       &    \\ 
 33276  &  1676  &   15    Ori          &   0.31  &  59.8  &   3.0 &       &       & a  \\ 
 37147  &  1905  &  122    \,$\tau$          &   0.23  &  109.9  &   5.5 & 1.77 & 0.07 &    \\ 
 37495  &  1935  &     \,$\nu$ 2Col          &   0.48  &  27.2  &   1.9 &       &       & a  \\ 
 37788  &  1955  &                      &   0.31  &  32.3  &   2.0 &       &       &    \\ 
 40292  &  2094  &                      &   0.29  &  38.1  &   2.1 &       &       &    \\ 
 41074  &  2132  &   39    Aur          &   0.35  &  87.7  &   4.4 & 1.78 & 0.06 &    \\ 
 46273  &  2384  &                      &   0.37  &  106.9  &   5.3 & 1.74 & 0.05 &    \\ 
 48737  &  2484  &   31\,$\xi$  Gem          &   0.44  &  66.0  &   3.3 & 1.78 & 0.04 &    \\ 
 49434  &  2514  &                      &   0.29  &  85.7  &   4.3 &       &       & a  \\ 
 50277  &  2551  &                      &   0.27  &  268.8  &  13.4 & 1.78 & 0.07 &    \\ 
 50635  &  2564  &   38    Gem          &   0.32  &  152.5  &   7.6 &       &       & b  \\ 
 51199  &  2590  &   19\,$\pi$  CMa          &   0.37  &  91.7  &   4.6 & 1.77 & 0.04 &    \\ 
 51733  &  2607  &                      &   0.39  &  102.3  &   5.1 &       &       & b  \\ 
 55052  &  2706  &   48    Gem          &   0.39  &  81.8  &   4.1 & 1.80 & 0.04 &    \\ 
 55892  &  2740  &                      &   0.32  &  50.7  &   2.5 & 1.93 & 0.10 &    \\ 
 56986  &  2777  &   55\,$\delta$ Gem          &   0.37  &  129.7  &   6.5 & 1.75 & 0.08 &    \\ 
 57749  &  2811  &                      &   0.35  &  43.4  &   6.3 &       &       &    \\ 
 58728  &  2846  &   63    Gem          &   0.45  &  30.9  &   2.2 &       &       & m3 \\ 
 58946  &  2852  &   62\,$\rho$ Gem          &   0.32  &  58.9  &   2.9 & 1.76 & 0.07 &    \\ 
 60111  &  2887  &    8\,$\delta$2CMi          &   0.32  &  117.5  &   5.9 & 1.80 & 0.05 & c  \\ 
 61110  &  2930  &   71$o$ Gem          &   0.41  &  91.1  &   4.6 & 1.78 & 0.04 &    \\ 
 62952  &  3015  &    4    Pup          &   0.34  &  127.5  &   6.4 & 1.80 & 0.08 & c  \\ 
 64379  &  3079  &                      &   0.47  &  42.6  &   2.6 & 1.56 & 0.05 & c  \\ 
 65925  &  3140  &                      &   0.40  &  71.6  &   3.6 &       &       & b  \\ 
 67483  &  3184  &   12    Cnc          &   0.48  &  52.3  &   4.1 & 1.50 & 0.07 & c  \\ 
 70958  &  3297  &    1    Hya          &   0.47  &  45.5  &   2.3 & 1.74 & 0.05 &    \\ 
 72943  &  3394  &                      &   0.33  &  56.8  &   2.8 & 1.70 & 0.06 & c  \\ 
 76143  &  3537  &                      &   0.42  &  83.0  &   4.2 & 1.79 & 0.03 &    \\ 
 76582  &  3565  &   63$o$2Cnc          &   0.20  &  90.5  &   4.5 & 1.80 & 0.05 &    \\ 
 77370  &  3598  &                      &   0.41  &  60.4  &   3.0 & 1.71 & 0.03 & c  \\ 
 79940  &  3684  &                      &   0.47  &  117.2  &   5.9 & 1.78 & 0.03 &    \\ 
 80671  &  3712  &                      &   0.41  &  26.2  &   2.1 &       &       & b  \\ 
 81937  &  3757  &   23    UMa          &   0.36  &  155.2  &   7.8 & 1.80 & 0.05 &    \\ 
 82434  &  3786  &     \,$\psi$ Vel          &   0.37  &  156.0  &  10.0 &       &       & b  \\ 
 82554  &  3795  &     \,$\iota$ Cha          &   0.45  &  129.7  &   6.5 & 1.63 & 0.05 &    \\ 
 83287  &  3829  &   42    Lyn          &   0.22  &  102.5  &   5.1 & 1.77 & 0.08 &    \\ 
 83962  &  3859  &                      &   0.41  &  140.3  &   7.0 & 1.72 & 0.08 &    \\ 
 84999  &  3888  &   29\,$\upsilon$ UMa          &   0.29  &  124.2  &   6.2 &       &       & s  \\ 
 87500  &  3969  &                      &   0.38  &  189.4  &   9.5 &       &       & b  \\ 
 88215  &  3991  &                      &   0.36  &  97.4  &   4.9 & 1.78 & 0.05 &    \\ 
 89025  &  4031  &   36\,$\zeta$ Leo          &   0.30  &  81.3  &   4.1 &       &       & b  \\ 
    \noalign{\smallskip}
    \hline
  \end{tabular}
\end{table*}

\addtocounter{table}{-1}
\begin{table*}
  \caption[]{continued}
  \begin{tabular}{rrcrrrrrc}
    \hline
    \hline
    \noalign{\smallskip}
    HD & HR & Name & $B-V^{\rm a}$ & $v\,\sin{i}$  & $¸\Delta~v\,\sin{i}$ & $q_{\rm 2}/q_{\rm 1}$ & $\Delta q_{\rm 2}/q_{\rm 1}$ & Shape \\
    & & & & [km\,s$^{-1}$] & [km\,s$^{-1}$] & & & Flag\\
    \noalign{\smallskip}
    \hline
    \noalign{\smallskip}
 89571  &  4062  &                      &   0.24  &  133.9  &   6.7  & 1.79 & 0.07  &    \\ 
 90089  &  4084  &                      &   0.39  &  56.2  &   2.8  & 1.70 & 0.09  &    \\ 
 90589  &  4102  &                      &   0.36  &  51.6  &   2.6  & 1.93 & 0.04  & c  \\ 
 91480  &  4141  &   37    UMa          &   0.34  &  36.9  &   4.0  &       &        &  a \\ 
 92787  &  4191  &                      &   0.32  &  58.2  &   2.9  & 1.84 & 0.10  & c  \\ 
 96202  &  4314  &     \,$\chi$1Hya          &   0.36  &  93.4  &   4.7  & 1.77 & 0.05  &    \\ 
 99329  &  4410  &   80    Leo          &   0.34  &  137.9  &   6.9  & 1.73 & 0.04  &    \\ 
 99453  &  4413  &                      &   0.49  &   4.9  &   1.2  &       &        & m2 \\ 
100623  &  4458  &                      &   0.81  &   4.0  &   2.2  &       &        & a  \\ 
103313  &  4555  &                      &   0.20  &  66.6  &   3.3  &       &        & s  \\ 
104731  &  4600  &                      &   0.41  &  11.0  &   2.0  &       &        &    \\ 
106022  &  4642  &                      &   0.42  &  77.2  &   3.9  & 1.86 & 0.06  &    \\ 
107192  &  4686  &                      &   0.34  &  68.1  &   3.4  &       &        & a  \\ 
107326  &  4694  &                      &   0.30  &  132.2  &   6.6  & 1.80 & 0.05  &    \\ 
109085  &  4775  &    8\,$\eta$ Crv          &   0.38  &  60.0  &   3.0  & 1.75 & 0.05  &    \\ 
109585  &  4797  &                      &   0.34  &  91.1  &   4.6  &       &        & b  \\ 
109799  &  4803  &                      &   0.33  &  34.0  &   6.0  &       &        &  a \\ 
110834  &  4843  &                      &   0.46  &  133.3  &   6.7  & 1.73 & 0.06  &    \\ 
111812  &  4883  &   31    Com          &   0.68  &  63.0  &   3.2  & 1.78 & 0.05  &    \\ 
113139  &  4931  &   78    UMa          &   0.36  &  91.7  &   4.6  &       &        & b  \\ 
114371  &  4966  &                      &   0.41  &   4.6  &   4.0  &       &        & m2 \\ 
114435  &  4970  &                      &   0.52  &  80.1  &   4.0  &       &        & b  \\ 
114837  &  4989  &                      &   0.48  &   8.8  &   3.0  &       &        &    \\ 
115810  &  5025  &                      &   0.27  &  99.1  &   5.0  & 1.82 & 0.06  & c  \\ 
116568  &  5050  &   66    Vir          &   0.41  &  37.0  &   1.9  &       &        &    \\ 
117360  &  5082  &                      &   0.45  &   4.0  &   2.5  &       &        &    \\ 
118261  &  5113  &                      &   0.49  &   4.7  &   3.6  &       &        & m3 \\ 
118889  &  5138  &                      &   0.34  &  140.6  &   7.0  & 1.82 & 0.06  & c  \\ 
119756  &  5168  &    1    Cen          &   0.39  &  63.9  &   3.2  & 1.78 & 0.03  &    \\ 
120987  &  5222  &                      &   0.44  &   8.5  &   2.5  &       &        &  a \\ 
121932  &  5253  &                      &   0.35  &  155.4  &   7.8  & 1.75 & 0.03  &    \\ 
122066  &  5257  &   48    Hya          &   0.48  &  41.1  &   2.1  & 1.84 & 0.03  &    \\ 
122797  &  5275  &                      &   0.40  &  33.6  &   5.0  &       &        &  a \\ 
123255  &  5290  &   95    Vir          &   0.34  &  157.8  &   7.9  & 1.81 & 0.05  & c  \\ 
124425  &  5317  &                      &   0.48  &  23.5  &   4.0  &       &        &    \\ 
124780  &  5337  &                      &   0.29  &  70.7  &   3.5  & 1.80 & 0.04  &    \\ 
125442  &  5364  &                      &   0.31  &  148.0  &  10.0  &       &        & b  \\ 
125451  &  5365  &   18    Boo          &   0.38  &  40.5  &   2.0  &       &        &    \\ 
127486  &  5426  &                      &   0.48  &  24.5  &   2.0  &       &        &  a \\ 
129153  &  5473  &                      &   0.23  &  105.7  &   5.3  & 1.78 & 0.06  & c  \\ 
129926  &  5497  &   54    Hya          &   0.31  &  112.5  &   5.6  & 1.74 & 0.03  & c  \\ 
132052  &  5570  &   16    Lib          &   0.31  &  113.2  &   5.7  & 1.77 & 0.06  &    \\ 
134083  &  5634  &   45    Boo          &   0.42  &  44.4  &   2.2  & 1.67 & 0.06  &    \\ 
137391  &  5733  &   51\,$\mu$ 1Boo          &   0.30  &  83.0  &   4.2  &       &        &  a \\ 
138917  &  5788  &   13\,$\delta$ Ser          &   0.26  &  85.8  &   5.4  & 1.99 & 0.06  & c  \\ 
139664  &  5825  &                      &   0.41  &  71.6  &   3.6  & 1.77 & 0.05  &    \\ 
141544  &  5882  &                      &   1.15  &   4.1  &   2.2  &       &        &    \\ 
142908  &  5936  &   12\,$\lambda$ CrB          &   0.35  &  75.7  &   3.8  & 1.80 & 0.05  & c  \\ 
143466  &  5960  &                      &   0.26  &  141.3  &   7.1  & 1.77 & 0.05  & c  \\ 
143790  &  5970  &                      &   0.47  &   5.5  &   3.0  &       &        &    \\ 
143928  &  5975  &                      &   0.40  &  23.7  &   1.2  &       &        &    \\ 
144415  &  5991  &                      &   0.29  &   7.8  &   1.7  &       &        & b  \\ 
146836  &  6077  &                      &   0.46  &  18.8  &   0.9  &       &        &    \\ 
147365  &  6091  &                      &   0.41  &  72.5  &   3.6  & 1.77 & 0.06  &    \\ 
147449  &  6093  &   50\,$\sigma$ Ser          &   0.33  &  77.1  &   3.9  & 1.80 & 0.03  &    \\ 
147787  &  6109  &     \,$\iota$ TrA          &   0.38  &  12.9  &  10.0  &       &        &  a \\ 
148048  &  6116  &   21\,$\eta$ UMi          &   0.39  &  84.8  &   4.2  & 1.78 & 0.05  &    \\ 
151613  &  6237  &                      &   0.37  &  47.4  &   2.4  & 1.84 & 0.07  &    \\ 
152598  &  6279  &   53    Her          &   0.31  &  60.7  &   3.0  &       &        & s  \\ 
153221  &  6300  &                      &   0.88  &   4.4  &   2.2  &       &        &    \\ 
    \noalign{\smallskip}
    \hline
  \end{tabular}
\end{table*}

\addtocounter{table}{-1}
\begin{table*}
  \caption[]{continued}
  \begin{tabular}{rrcrrrrrc}
    \hline
    \hline
    \noalign{\smallskip}
    HD & HR & Name & $B-V^{\rm a}$ & $v\,\sin{i}$  & $¸\Delta~v\,\sin{i}$ & $q_{\rm 2}/q_{\rm 1}$ & $\Delta q_{\rm 2}/q_{\rm 1}$ & Shape \\
    & & & & [km\,s$^{-1}$] & [km\,s$^{-1}$] & & & Flag\\
    \noalign{\smallskip}
    \hline
    \noalign{\smallskip}
153580  &  6314  &     \,$\epsilon$2Ara          &   0.49  &  45.4  &   2.3  &       &        & b  \\ 
156098  &  6409  &                      &   0.51  &   5.8  &   3.2  &       &        &    \\ 
156295  &  6421  &                      &   0.21  &  107.4  &   5.4  & 1.79 & 0.07  &    \\ 
157728  &  6480  &   73    Her          &   0.22  &  73.0  &   3.7  &       &        & s  \\ 
158741  &  6522  &                      &   0.36  &  147.5  &   7.4  &       &        & s  \\ 
164259  &  6710  &   57\,$\zeta$ Ser          &   0.39  &  69.3  &   3.5  & 1.75 & 0.04  &    \\ 
171834  &  6987  &                      &   0.38  &  71.3  &   3.6  & 1.77 & 0.04  &    \\ 
175813  &  7152  &     \,$\epsilon$ CrA          &   0.39  &  148.5  &   8.0  & 1.97 & 0.15  &    \\ 
177171  &  7213  &     \,$\rho$ Tel          &   0.53  &  68.5  &   4.0  &       &        & m2 \\ 
180868  &  7315  &   25\,$\omega$1Aql          &   0.19  &  83.0  &   4.2  &       &        & s  \\ 
182640  &  7377  &   30\,$\delta$ Aql          &   0.31  &  87.3  &   4.4  & 1.67 & 0.06  & c  \\ 
185124  &  7460  &   42    Aql          &   0.42  &  87.0  &   4.4  & 1.71 & 0.03  &    \\ 
186005  &  7489  &   55    Sgr          &   0.31  &  149.9  &   7.5  & 1.79 & 0.04  &    \\ 
187532  &  7553  &   51    Aql          &   0.40  &  77.5  &   3.9  & 1.75 & 0.03  &    \\ 
189245  &  7631  &                      &   0.49  &  72.6  &   3.6  & 1.74 & 0.03  &    \\ 
    \noalign{\smallskip}
    \hline
  \end{tabular}
  \begin{list}{}{}
  \item[$^{\rm a}$] \citet{BSC}
  \end{list}
\end{table*}

%%% Local Variables: 
%%% mode: latex 
%%% TeX-master: "master" 
%%% End:


\begin{thebibliography}{} 
  
\bibitem[Donati \& Collier Cameron, 1997]{Donati97} Donati J.-F.,
  Collier Cameron A., 1997, MNRAS 291, 1
  
\bibitem[G\l\c{e}bocki et al., 2000]{Glebocki00} G\l\c{e}bocki R., \&
  Stawikowski A. 2000, Acta Astron., 50, 509
  
\bibitem[Hauck \& Mermilliod, 1998]{Hauck98} Hauck B., \& Mermilliod
  M., 1998, A\&AS, 129, 431
  
\bibitem[Hoffleit \& Warren, 1991]{BSC} Hoffleit E.D., \& Warren Jr.
  W.H., 1991, The Bright Star Catalogue, 5th Revised Ed.
  
\bibitem[Kupka et al., 1999]{VALD} Kupka, F., Piskunov, N.E.,
  Ryabchikova, T.A., Stempels, H.C., \& Weiss, W.W., 1999, A\&AS, 138,
  119
  
\bibitem[Moon, 1985]{Moon85a} Moon T.T., 1985, Comm.  Univ. London
  Obs. 78
 
\bibitem[Moon \& Dworetsky, 1985]{Moon85b} Moon T.T., \& Dworetsky
  M.M., 1985, MNRAS, 217, 305
 
\bibitem[Napiwotzki et al., 1993]{Napiwotzki93} Napiwotzki R.,
  Sch\"onberger D., \& Wenske V., 1993, A\&A, 268, 653
 
\bibitem[Reiners \& Schmitt, 2002]{Reiners02} Reiners, A., \& Schmitt,
  J.H.M.M., 2002, A\&A, 384, 555
  
\bibitem[Reiners \& Schmitt, 2003]{Reiners03} Reiners, A., \& Schmitt,
  J.H.M.M., 2003, A\&A, 398, 647
  
\bibitem[Reiners, 2003]{Reiners03a} Reiners, A., 2003, accepted by
  A\&A
 
\bibitem[Royer et al., 2002]{Royer02} Royer F., Grenier S., Baylac
  M.-O., Gomez A.E., \& Zorec J., 2002, A\&A 393, 897
  
\bibitem[Siess et al., 2000]{Siess00} Siess L., Dufour E., \&
  Forestini M., 2000, A\&A, 358, 593
 
\bibitem[Uesugi \& Fukuda, 1982]{Uesugi82} Uesugi A., \& Fukuda I.,
  1982, Revised Catalogue of Rotational Velocities, Department of
  Astronomy, Kyito Univ., Japan

\end{thebibliography}
\end{document}